\documentclass[pre,aps,twocolumn,showpacs,floatfix]{revtex4-1}
\usepackage{graphicx,bm,amsmath,dcolumn,longtable}
\usepackage{geometry}
\graphicspath{{figures}}
\begin{document}
\newcommand{\half}{\frac{1}{2}}
\newcommand{\ith}{^{(i)}}
\newcommand{\im}{^{(i-1)}}
\newcommand{\gae}
{\,\hbox{\lower0.5ex\hbox{$\sim$}\llap{\raise0.5ex\hbox{$>$}}}\,}
\newcommand{\lae}
{\,\hbox{\lower0.5ex\hbox{$\sim$}\llap{\raise0.5ex\hbox{$<$}}}\,}
\newcommand{\be}{\begin{equation}}
\newcommand{\ee}{\end{equation}}

\title{Exact critical points of the O($n$) loop model on the martini and the 3-12 lattices}
\author{Chengxiang Ding}  
\affiliation{Physics Department, Anhui University of Technology, Maanshan 243002, P. R. China}
\author{Zhe Fu}
\affiliation{Physics Department, Beijing Normal University,
Beijing 100875, P. R. China}
\author{Wenan Guo}
\email[Corresponding author: ]{waguo@bnu.edu.cn}  
\affiliation{Physics Department, Beijing Normal University,
Beijing 100875, P. R. China}

\date{\today} 
\begin{abstract}
We derive the exact critical line of the O($n$) loop model on the martini lattice as a function of the loop weight $n$. 
A finite-size scaling analysis based on transfer matrix calculations is also performed.
The numerical results coincide with the theoretical predictions with an accuracy up to 9 decimal places.
In the limit $n\to 0$, this gives the exact connective constant $\mu=1.7505645579\cdots$ of self-avoiding walks on the 
martini lattice.
Using similar numerical methods, we also study the O($n$) loop model on the 3-12 lattice. We obtain similarly precise agreement with the exact critical points 
given by Batchelor [J. Stat. Phys. {\bf 92}, 1203 (1998)].
\end{abstract}
\maketitle 
{\it Introduction.}
The O($n$) loop model \cite{on1} originates from the high-temperature expansion of the O($n$) spin model \cite{on2}. It can be considered a model describing a nonintersecting loop gas.
On lattices with coordination number three, the partition function is very simple:
\begin{equation}
Z=\sum\limits_{\cal G}x^{b}n^{l},
\label{Zloop}
\end{equation}
where the sum is over all configurations of non-intersecting loops denoted as $\cal G$; 
$x$ is the weight of a bond (an edge occupied by loop segments), or an occupied vertex, and $n$ the weight of a
loop.  $b$ is the number of bonds or occupied vertices, and $l$ the number of loops. 

Generally speaking, there is a high-temperature phase with dilute loops and a low-temperature phase with dense loops. At the transition point $x_c$, the longest loop grows to infinity and begins to percolate the system.  The critical properties of this transition  are 
universal, which are well described by the Coulomb gas theory \cite{cg}. 
However, the determination of the critical points of this model on various lattices remains 
to be treated case by case. Exact O($n$) critical lines have been 
found on the honeycomb lattice \cite{onhoneycomb, baxteron, hycONhenk}, the square lattice \cite{onsquare}  and the triangular lattice \cite{ontriangle}. 
In addition to this transition point, there are several other branches of critical 
behavior, e.g., `branch 0', which describes a higher critical point, 
as reported in Refs.  \cite{ontriangle,on,onkagome}. 
 
In the $n \to 0$ limit, the critical O($n$) loop model describes long polymers in a good solvent or 
self-avoiding walks (SAWs)\cite{deGennes}.  The 
study of the O($n$) loop model has led to a wealth of information on the configuration properties of SAWs
\cite{sawrev}. The number of configurations of SAWs in $k$ steps, i.e., $C_k$ scales as 
\cite{sawu}
\begin{eqnarray}
C_k \sim \mu^k k^{\gamma-1}
\end{eqnarray}
for large $k$. $\gamma=43/32$ is a universal critical exponent,  which can be
obtained via the Coulomb gas theory \cite{onhoneycomb}.
$\mu$ is the connective constant which is lattice dependent, and equals to $1/x_c$ of the $n \to 0$ loop model.
Although the studies on the SAWs have advanced a lot since it was introduced\cite{sawfirst}, the values of $\mu$ for most of two-dimensional lattices are  found numerically\cite{sea,sawhon,sawhon3,sawsemi, sawhon2}. 
But the exactly known critical line of the O($n$) loop model on the honeycomb lattice  \cite{onhoneycomb}
provides the exact result $\mu=\sqrt{2+\sqrt{2}}$ for the model on the honeycomb lattice.
The critical line of  the honeycomb O($n$) loop model was found by an exact
mapping on a Potts model
\cite{onhoneycomb} and by using the Bethe Ansatz\cite {baxteron, hycONhenk}:
\begin{equation}
\label{xc}
x_c^2=\frac{1}{2+\sqrt{2-n}}.
\end{equation}
The phase diagram inferred from this result has been well verified by different numerical methods\cite{hycONTM, honeycombMC}.
Batchelor derived  the critical points of the O($n$) loop model on the 3-12 lattice by mapping the honeycomb
loop model to the 3-12 lattice\cite{on312}:
\begin{equation}
(\frac{x_c^2+x_c^3}{1+x_c^3n})^2=\frac{1}{2+\sqrt{2-n}}. \label{xc312}
\end{equation}
This result is in agreement with existing exact results \cite{on312} for the cases $n=1$ (Ising) and $n=0$. 
Making use of this result, Batchelor obtained $\mu=1.711 041 \cdots$ for the 3-12 lattice, 
which coincides with the result previously found by Jensen and Guttmann\cite{sawsemi} using other methods.
In the present paper we shall provide some independent numerical results for
the critical point of the O($n$) loop model on the 3-12 lattice, and on its phase behavior.

Inspired by Batchelor's work, we studied the O($n$) loop model on the martini lattice. This lattice
was first proposed by Scullard\cite{scullard} in the study of percolation.  The percolation threshold 
and the critical points of the $q$-state Potts model\cite{potts,wfypotts} on this lattice are known 
exactly\cite{qAC}, but the critical points of the O($n$) loop model are not known yet.
In this paper, we derive the exact critical points of the O($n$) loop model as a function of $n$ on the martini lattice.
In the limit $n \to 0$, the result gives the exact connective constant of
SAWs on the martini lattice.
In addition, we build the transfer matrix (TM) and apply a finite-size scaling analysis for a numerical study of the model on the martini and the 3-12 lattice.

{\it Critical points of O($n$) loop model on the martini lattice.}
\begin{figure}[htpb]
\includegraphics[scale=0.20]{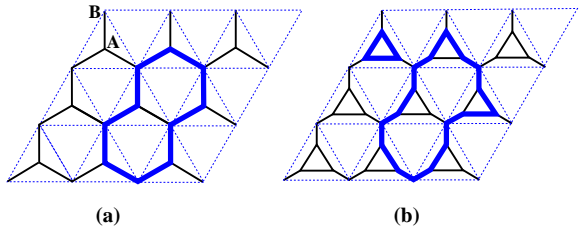}
\caption{(Color online) The O($n$) loop model on the honeycomb lattice (a) and the martini lattice (b).}
\label{cfg2}
\end{figure}
\begin{figure}[htpb]
\includegraphics[scale=0.15]{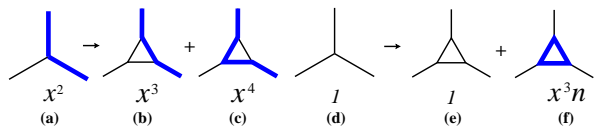}
\caption{(Color online) Mapping of the vertex configurations. The weights of the different
vertices are shown. The same weights apply to versions rotated by $\pm 2 \pi/3$.}
\label{map2}
\end{figure}
Consider a honeycomb lattice with two sublattices $A$ and $B$.
A loop configuration $\cal G$ also denotes the occupations of sublattice $A$, as
shown in Fig. \ref{cfg2}(a).  
We rewrite the partition function (\ref{Zloop}) 
in the following way:
\begin{eqnarray}
Z_{h}&=&\sum\limits_{\cal G} x^{v} n^l
=\sum\limits_{\cal {\cal G}} (x^2)^{v_A} 1^{V_A-v_A} n^l, 
\label{zhoney1}
\end{eqnarray}
where $v=b$ is the number of vertices visited by loop segments,  
$V_A$ and $v_A=v/2$ are the number of vertices and the number of visited 
vertices of sublattice $A$, respectively. 
Thus the weight of a visited $A$ vertex shown in Fig. \ref{map2}(a) is $x^2$,
the weight of an empty $A$ vertex shown in Fig. \ref{map2}(d) is 1. 

Now consider the O($n$) loop model on the martini lattice, 
which is constructed by replacing the `star' (an $A$ vertex) shown in Fig. \ref{map2}(d) 
by the structure shown in Fig. \ref{map2}(e). 
Each occupied (empty) $A$ vertex corresponds to two possible occupations 
on that structure, as shown in Fig. \ref{map2} (b) and (c)
((e) and (f)).
Thus, any given configuration $\cal G$ of loops on the honeycomb lattice 
maps to the sum of $2^{V_A}$ possible loop configurations on the martini lattice.  
Let $x$ be the weight of a bond for the martini loop model, the partition 
function of the loop model on the 
martini lattice can be obtained by summing on $\cal G$: 
\begin{eqnarray}
Z_{m}
 &=&(1+x^3n)^{V_A} \sum\limits_{\cal G} (\frac{x^3+x^4}{1+x^3n})^{v_A}n^l. \label{zmar1}
\end{eqnarray}
Mapping 
$(x^3+x^4)/(1+x^3n) \to x^2$, we obtain $Z_h$ in (\ref{zhoney1}) multiplied by a trivial factor.
It follows the critical points $x_c$ of the O($n$) loop model on the martini lattice 
\begin{eqnarray}
\frac{{x_c}^3+{x_c}^4}{1+{x_c}^3n}=\frac{1}{2+\sqrt{2-n}}.\label{xcmar}
\end{eqnarray}

This result is in agreement with an existing result for 
the O(1) loop model,  which is equivalent to the high temperature expansion of the Ising model model \cite{baxter} 
with 
$x={\rm tanh}K^I$,
where $K^I$ is the coupling of Ising spins sitting on the vertices of the martini lattice. 
According to (\ref{xcmar}), 
$K^I_c=0.749790959036 \cdots$, 
which coincides with the exactly known critical point of the $q=2$ Potts model on the martini lattice \cite{qAC}.

As another special case, we present the exact connective constant of the SAW on the martini lattice. 
The substitution $\mu=1/x_c$ in (\ref{xcmar}) for $n=0$ determines $\mu$ as the solution of  
\begin{eqnarray}
\frac{1}{\mu^3}+\frac{1}{\mu^4}=1-\frac{\sqrt{2}}{2},
\end{eqnarray}
which yields $\mu=1.750564567897\cdots$. 

We may further generalize above results by allowing
bonds on the small triangles in Fig. \ref{map2}(e) have weight ($x_t$) different from
those on the remaining ones ($x_s$). 
Following the mapping described above, we thus obtain a critical line in the $x_t$
versus $x_s$ plane for a given $n$:
\be
\frac{{x_s}^2 x_t+{x_t}^2 {x_s}^2}{1+{x_t}^3n}=\frac{1}{2+\sqrt{2-n}}.\label{xcmar1}
\ee
For the 3-12 lattice, the critical line of the generalized model is
\begin{eqnarray}
\frac{x_s(x_t+x_t^2)}{1+x_t^3n}=\frac{1}{\sqrt{2+\sqrt{2-n}}}. \label{312gen}
\end{eqnarray}

{\it Finite-size scaling and transfer matrix calculation.}
Consider the lattice (the 3-12 or the martini lattice) wrapped on a cylinder with circumference $L$. 
The magnetic correlation function of the O($n$) spin model is translated as the probability that 
two sites at a distance $r$ are linked by a single loop segment \cite{cg},
$g_r=Z^{\prime}/Z$, where $Z^{\prime}=\sum_{{\cal G}^{\prime}} x^bn^l$, and ${\cal G}^{\prime}$ denotes the 
configurations that
connect sites 0 and $r$ by precisely one single loop segment. In our transfer-matrix analysis of the finite-size-scaling
behavior, it is sufficient to substitute the configurations ${\cal G}^{\prime}$ that connect any site of row 0
to any site of a row at a distance $r$ as measured in the length direction of the cylinder. The exponential
decay of $g_r$ at large distances is determined by the magnetic gap in the eigenvalue spectrum of the transfer matrix.
The scaled magnetic gap  
$X_h(x,L)=\frac{L \zeta}{2\pi} \ln ({\Lambda^{(0)}}/{\Lambda^{(1)}}),$
where $\Lambda^{(0)}$ and $\Lambda^{(1)}$ is the largest eigenvalue of the TM  
for $Z$ and $Z^{\prime}$, respectively. 
$\zeta$ is a geometrical factor determined by the ratio between the unit of $L$ and the thickness of a row 
added by the transfer matrix. Another scaled gap $X_t(x,L)=\frac{L \zeta}{2\pi} \ln ({\Lambda^{(0)}}/{\Lambda^{(2)}})$,
describes the exponential decay of the energy-energy correlation, with $\Lambda^{(2)}$ the
second eigenvalue of the TM for $Z$.

The TM techniques 
of the O($n$) loop model are well described in the literature, e.g., see \cite{on}.
The procedure of sparse matrix decomposition for the martini and the 3-12 lattice
equals to that for the honeycomb lattice with the adding units suitably chosen 
\cite{hycONTM}.  For further details, see \cite{pottsTM,TM1}.

\begin{figure}[htbp]
\includegraphics[scale=0.73]{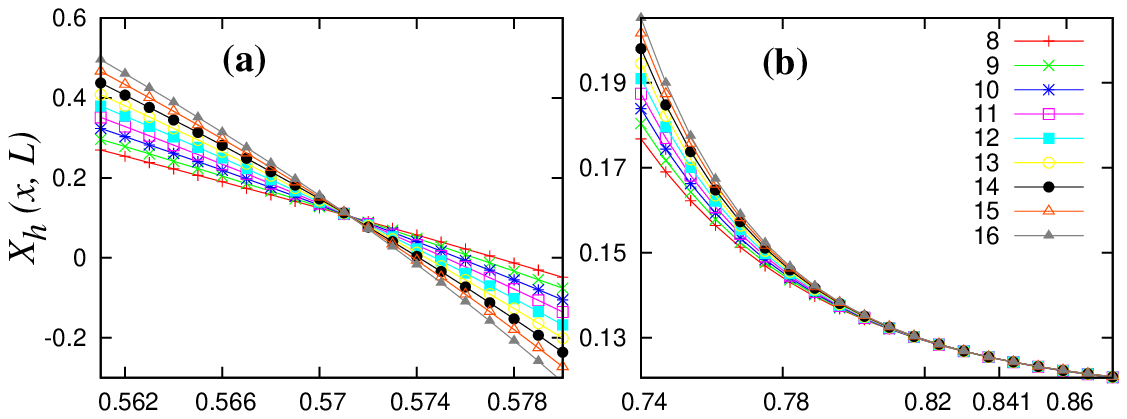}
\includegraphics[scale=0.73]{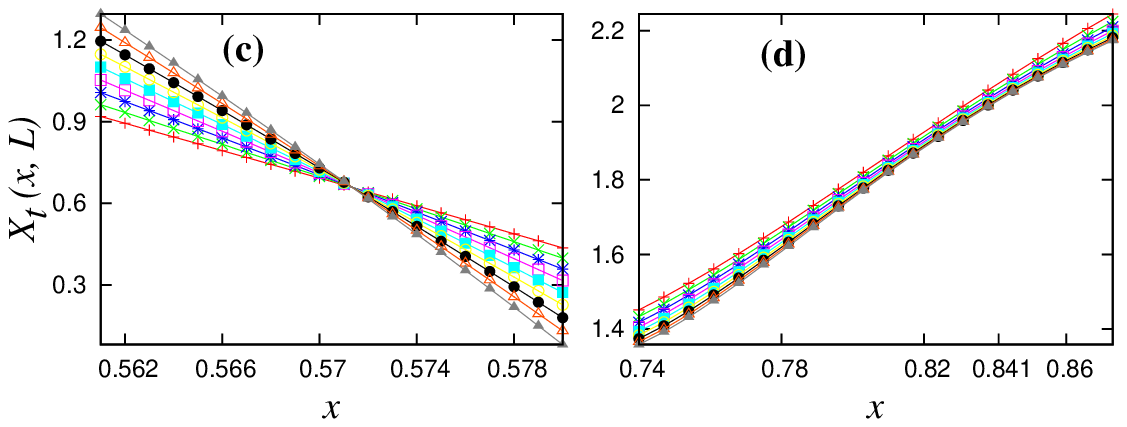}
\caption{(Color online) Scaled gaps $X_h(x,L), X_t(x,L)$ versus $x$ for the O($n$) loop model on the martini lattice with $L$=8 to 16. (a) and (c): $n=0$; 
(b) and (d): $n=2$, with $x_c=0.840896$ indicated.
Lines connecting data points are added to guide the eye. }
\label{xhc2}
\end{figure}

According to the finite-size scaling \cite{fss} and the conformal invariance \cite{conformal} 
theory, the scaled gap 
$X_i(x,L)$, in the vicinity of the critical point,  satisfies
\begin{eqnarray}
\label{xh} 
X_i(x,L)=X_i+a(x-x_c)L^{y_t}+buL^{y_u}+\cdots,
\end{eqnarray}
where $X_i (i=h,t)$ is the magnetic and the temperature scaling dimension, respectively; $y_t$ is the 
thermal exponent; $u$ denotes the leading irrelevant field,
$y_u$ is the associated irrelevant exponent.  $a, b$ are unknown constants. 
Such behavior is illustrated in Fig. \ref{xhc2}(a) and (c) for the case $n=0$.
The critical point is estimated by numerically solving $x$ in the equation
$X_h(x,L)=X_h(x,L-1)$,
with system sizes up to $L=16$.  The solution $x_c(L)$ converges to the critical value $x_c$ as 
\begin{eqnarray}
\label{xceq}
x_c(L)=x_c+a^{\prime}uL^{y_u-y_t}+\cdots,
\end{eqnarray}
where $a^{\prime}$ is an unknown constant.
The numerical estimations and the theoretical predictions of the critical points of the martini lattice are 
listed in Table \ref{tabmartini}. 
Our numerical estimations coincide with the theoretical predictions in $9$ decimal numbers. 

\begin{ruledtabular}
\begin{table*}[htbp]
\caption{ Critical points $x_c$, conformal anomaly $c$,  magnetic and temperature scaling dimensions $X_h, X_t$
of the two-dimensional O($n$) loop model on the martini lattice. (T=Theoretical prediction, N=Numerical estimation.)}
\begin{tabular}{lllllllll}
      $n$ &$x_c$(T)     &  $x_c$(N)     & $c$(T)  & $c$(N)      & $X_h$(T)&$X_h$(N)     &$X_t$(T)& $X_t$(N)\\
    \hline
    0     &0.571244285  &0.571244285(1) &     0   &     0       &0.1041667&0.104166(1)  &2/3       &0.666668(2)\\
    0.25  &0.584248605  &0.584248605(1) &0.1300704&0.1300705(2) &0.1100192&0.1100193(1) &0.7395254  &0.739526(1)\\
    0.5   &0.598867666  &0.59886768(2)  &0.2559499&0.255950(1)  &0.1154420&0.1154420(1) &0.8177559  &0.817756(1)\\
    0.75  &0.615559079  &0.6155590(1)   &0.3788781&0.3788783(3) &0.1204452&0.1204452(1) &0.9035105 &0.9035104(2)\\
    1.0   &0.635024224  &0.63502422(1)  &0.5      &0.500000(1)  &0.125    &0.12500000(1)&1         &1.00000(1)\\
    1.25  &0.658437850  &0.65843786(1)  &0.6205051&0.6205053(2) &0.1290128&0.1290127(1) &1.1126008 &1.1126007(1)\\
    1.5   &0.688067393  &0.68806739(2)  &0.7418425&0.741842(1)  &0.1322435&0.1322434(1) &1.2518912 &1.25189(1)\\
    1.75  &0.729662053  &0.729664(3)    &0.8662562&0.8662563(1) &0.1339623&0.13396(1)   &1.4457176 &1.445718(1)\\
    2.0   &0.840896415  & -             &1        &1.000000(1)  &0.125    &0.1250000(1) &2         &2.00000(1)\\
\end{tabular}
\label{tabmartini}
\end{table*}
\end{ruledtabular}

The universal values of $X_i$ and the conformal anomaly $c$ of the two-dimensional O($n$) loop model are exactly known as
\cite{cg,conformal} 
\begin{eqnarray}
\label{cxh}
c&=&1-\frac{6(g-1)^2}{g}, \nonumber \\
X_h&=& 1 -\frac{1}{2g}-\frac{3g}{8}, \ \  X_t=\frac{4-2g}{g},
\end{eqnarray}
where $n=-2\cos(\pi g)$, $1\le g \le 2$.

At criticality,  $X_i(L)$  converges as follows to $X_i$ with increasing system size $L$  
\begin{eqnarray}
\label{xhc}
X_i(L)=X_i+b^\prime L^{y_u}+\cdots,
\end{eqnarray}
where $b^\prime$ is an unknown constant. 
The free energy density $f(L)=\zeta \ln \Lambda^{(0)}_L/L$ scales as \cite{BCN, Affleck}
\begin{eqnarray}
\label{fc}
f(L)=f(\infty)+\frac{\pi c}{6L^2}.
\end{eqnarray}
We then calculate $f(L),~X_t(L)$ and $X_h(L)$ for a sequence of systems up to size $L=16$ at the critical 
points (\ref{xcmar}). 
Fitting the data according to (\ref{xhc}) and (\ref{fc}), 
we obtain the scaling dimensions $X_h,~X_t$ and the conformal anomaly $c$, which are also
listed in Table \ref{tabmartini}.  Our numerical estimations are in agreement with the theoretical 
predictions with a high accuracy. For $X_h, X_t$, our results are also consistent with the Monte Carlo results
for $n \ge 1$ \cite{honeycombMC, ONpercolation}.

When $n$ approaches 2, $y_t=2-X_t\rightarrow 0$. 
The corrections to scaling due to the leading irrelevant field become relatively strong.
At $n=2$, $y_t$ is exactly $0$, so that intersecting points between the
curves $X_i(x, L)$ and $X_i(x, L-1)$ may be absent, as already suggested by 
(\ref{xh}), and as indeed observed 
in Fig. \ref{xhc2}(b) and (d). Therefore, we can't numerically determine the critical point 
$x_c$ in the usual way. 
However, $c$ and $X_h, X_t$ are still estimated at the theoretical critical point. 

Similar analysis is also performed to the 3-12 lattice.
Theoretical predictions and numerical estimations of critical points for several values of $n$, which
agree in a high accuracy,  are listed in Table \ref{tab312}. 
>From Tables \ref{tabmartini} and \ref{tab312}, we can see that the values of $X_h, X_t$ and $c$ for a two-dimensional 
O($n$) loop model on 
the martini lattice coincide with those of the 3-12 lattice, as expected by the hypothesis of universality.
\begin{ruledtabular}
\begin{table*}[t]
\caption{ Critical points $x_c$, conformal anomaly $c$,  magnetic and temperature scaling dimensions $X_h, X_t$
of the two-dimensional O($n$) loop model on the 3-12 lattice. (T=Theoretical prediction, N=Numerical estimation.)}
\begin{tabular}{lllllllll}
     $n$&  $x_c$(T) &  $x_c$(N)    & $c$(T)  & $c$(N)      & $X_h$(T)& $X_h$(N)   &$X_t$(T)&$X_t$(N) \\
    \hline
    0   &0.584439429&0.584439429(1)&0        &0            &0.1041667&0.104167(1) &2/3       &0.666668(1)\\
    0.25&0.601034092&0.601034092(1)&0.1300704&0.1300705(2) &0.1100192&0.1100193(1)&0.7395254 &0.739526(1) \\
    0.5 &0.620240607&0.62024060(1) &0.2559499&0.255950(3)  &0.1154420&0.115442(1) &0.8177559 &0.817756(1)\\
    0.75&0.642967899&0.6429678(1)  &0.3788781&0.3788783(2) &0.1204452&0.1204452(1)&0.9035105 &0.903510\\
    1.0 &0.670697664&0.67069766(1) &0.5      &0.499999(1)  &0.125    &0.1250000(1)&1         &1.000000(1)\\
    1.25&0.706102901&0.70610291(1) &0.6205051&0.620505(1)  &0.1290128&0.1290127(1)&1.1126008 &1.112600(1)\\
    1.5 &0.754845016&0.754845(1)   &0.7418425&0.741842(1)  &0.1322435&0.1322435(1)&1.2518912 &1.25189(1)\\
    1.75&0.833205232&0.83320(2)    &0.8662562&0.86626(1)   &0.1339623&0.13396(1)  &1.4457176 &1.445718(2)\\
    2.0 &1.172534677&-             &1        &0.999999(1)  &0.125    &0.1250000(1)&2         &2.00000(1)\\
\end{tabular}
\label{tab312}
\end{table*}
\end{ruledtabular}

{\it Conclusion.}
We derived the exact critical points of the O($n$) loop model on the martini lattice, and
performed a finite-size scaling analysis based on numerical TM calculations.  
Our numerical estimations agree with the theoretical predictions, within a margin that is typically of order
$10^{-9}$. This rather high precision may be related to the vanishing of the leading irrelevant field in
the Nienhuis result \cite{onhoneycomb} for the critical line.

In the limit $n \to 0$, the critical point gives the exact connective constant $\mu=1.7505....$ of the SAWs on the 
martini lattice.

The exact critical points of the O($n$) loop model on the 3-12 lattice derived by Batchelor are 
also verified.

The conformal anomaly, the magnetic and  the temperature scaling dimensions of the O($n$) models on the two lattices 
are numerically calculated. The estimations coincide with the theoretical predictions, 
as expected according to the universality hypothesis.

{\it Acknowledgment.}
We thank Prof. F. Y. Wu and Prof. H. W. J. Bl\"ote for a critical reading of
the manuscript and valuable suggestions.
This work is supported by the NSFC under Grant No.~11175018, 
the NCET-08-0053 and 
the HSCC of Beijing Normal University.

\end{document}